\documentclass[10pt,conference,a4paper]{IEEEtran}
\usepackage{times,multicol,amsmath,epsfig}

\title{Informed stego-systems in active warden context: statistical
undetectability and capacity }
\author{%
{Sofiane Braci{\small $~^{\#1}$}, Claude Delpha{\small $~^{\#1}$},
R\'emy Boyer{\small $~^{\#1}$}, Ga\"etan Le Guelvouit{\small $~^{{*} 2}$}}%
\vspace{1.6mm}\\
\fontsize{10}{10}\selectfont\itshape
$^{\#}$\,Laboratoire des Signaux et Syst\'emes (L2S) \\ CNRS, Universit\'e Paris-Sud XI (UPS), SUPELEC\\
 \fontsize{9}{9}\selectfont\ttfamily\upshape
$^{1}$\,$\{$sofiane.braci,claude.delpha,remy.boyer$\}$@lss.supelec.fr\\
\vspace{1.2mm}\\
\fontsize{10}{10}\selectfont\rmfamily\itshape

$^{*}$France T\'el\'ecom R\&D -- Orange Labs\\
4, rue du Clos Courtel -- 35512 Cesson-S\'evign\'e Cedex \\
\fontsize{9}{9}\selectfont\ttfamily\upshape
$^{2}$\,gaetan.leguelvouit@orange-ftgroup.com }
    
\begin{document}

            \maketitle

            \begin{abstract}
Several authors have studied stego-systems based on Costa scheme, but just a few ones gave both theoretical and experimental justifications of these schemes performance in an active warden context. We provide in this paper a steganographic and comparative study of three informed stego-systems in active warden context: scalar Costa scheme, trellis-coded quantization and spread transform scalar Costa scheme. By leading on analytical formulations and on experimental evaluations, we show the advantages and limits of each scheme in term of statistical undetectability and capacity in the case of active warden. Such as the undetectability is given by the distance between the stego-signal and the cover distance. It is measured by the Kullback-Leibler distance.
            \end{abstract}

%
            \section*{Introduction} \label{sec:intro}

In data hiding, a very old field named steganography is used since
the Antiquity. As defined by Cox~\emph{et al.}~\cite{label_ref_cox},
steganography denotes ``\emph{the practice of undetectability
altering a work to embed a message}". In the classical problem of
the prisoners~\cite{simmons84}, Alice and Bob are in prison and try
to escape. They can exchange documents, but these documents are
controlled by an active warden named Wendy. Cox~\cite{label_ref_cox}
defines the warden as active when ``\emph{she intentionally
modifies the content sent by Alice prior to receipt by Bob}".
These modifications can slightly modify the content and
degrade the hidden information. In this work, we consider that all
modifications performed by Wendy are modeled by an Additive White
Gaussian Noise (AWGN) and we propose to study the limits of such
systems. Since our specific active warden context is similar to the
case of watermarking with AWGN channel, we propose to study the
capacity according to the Shannon definition~\cite{label_ref_cox} as
the maximum information bits that can be embedded in one sample
subject to certain level of the active warden attack (an AWGN attack
in this case). In sequel, we evaluate the statistical
undetectability by the Kullback-Leibler Distance (KLD) between
the probability density functions (p.d.f.) of the stego-signal and
the cover-signal, since the warden detects the message by comparing
the stego-document probability density function with that of the
cover-document. In~\cite{label_ref_Cachin}, author used KLD
to evaluate the security of stego-systems in the context of the
passive warden. In this work, Cachin's security criterion is not
used since the context is different (active warden context).

We propose here to base our comparative study on informed data
hiding schemes as the Scalar costa scheme (SCS). One of the major
work already proposed on these type of scheme by Guillon~\emph{et al.}~\cite{guillon02} experimentally found that SCS is statistically
detectable due to artifacts in the p.d.f. of the stego-signal. The
way proposed to make it undetectable is the use of a specific
compressor on the signal leads to a less flexible scheme. Le~Guelvouit~\cite{label_ref_Gaetan} proposed to use Trellis-Coded
Quantization (TCQ) in order to hide the message: the author shows
experimentally that the p.d.f. of the stego-signal is not affected
by the embedded message. We fully complete this study and also
theoretically demonstrate this result. Moreover, we propose in this
work an evaluation of steganographic performance in an active warden
context of the Spread Transform Scalar Costa Scheme (ST-SCS)~\cite{eggers03}, which is often use for robust watermarking. We
demonstrate with experiments and analytic formulations the good
statistical undetectability level of this system, then we compare
its capacity and the compromise between the capacity and the
statistical undetectability with other systems. \\

Let us first list some notational conventions used in this paper.
Vectors are notes in bold font and sets in black board font. Data
are written in small letters, and random variables in capital ones;
$\textbf{s}[i]$ is the $i^{\scriptsize \textrm{th}}$ component of
vector $\textbf{s}$. The probability density function of random
variable $S$ is denoted by $p_S(.)$.

       \section{Analysis of scalar Costa scheme}

Eggers~\emph{et al.}~\cite{eggers03} have introduced a sub-optimal
scheme based on the Costa's ideas~\cite{costa83}. The authors
propose to construct a codebook from the reconstruction points of a
scalar quantizer. This approach is called \emph{Scalar Costa Scheme}
(SCS) and has a high capacity for optimal value of Costa's factor
$\alpha$. However, it has been shown~\cite{label_ref_Gaetan} that the regular partitioning of scalar
quantizers generates many artifacts in the p.d.f. of the marked signal.

\begin{figure}
        \begin{center}
            \begin{tabular}{cc}
                \includegraphics[width=0.22\textwidth , height=3.5cm]{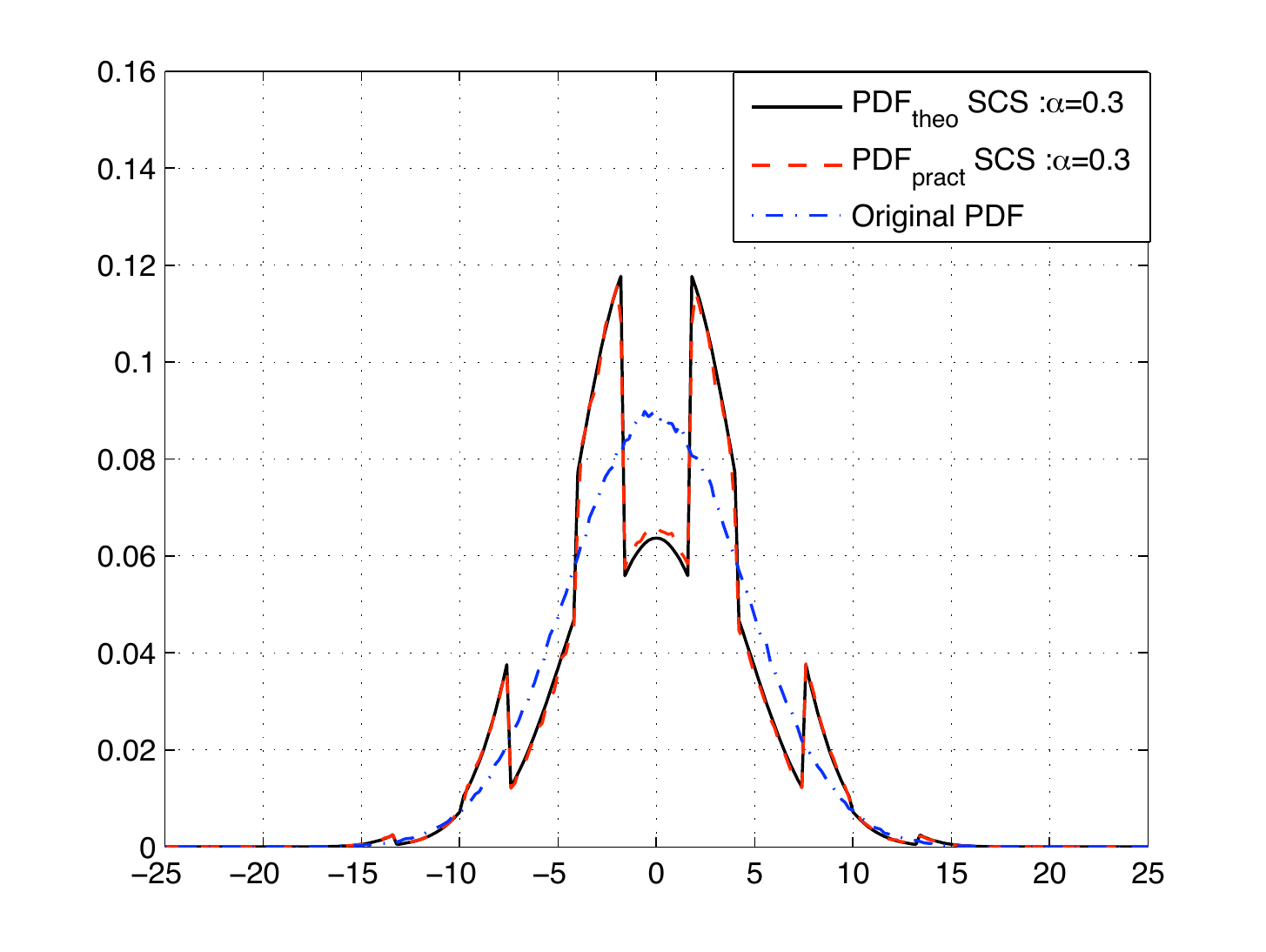} &
                \includegraphics[width=0.22\textwidth , height=3.5cm]{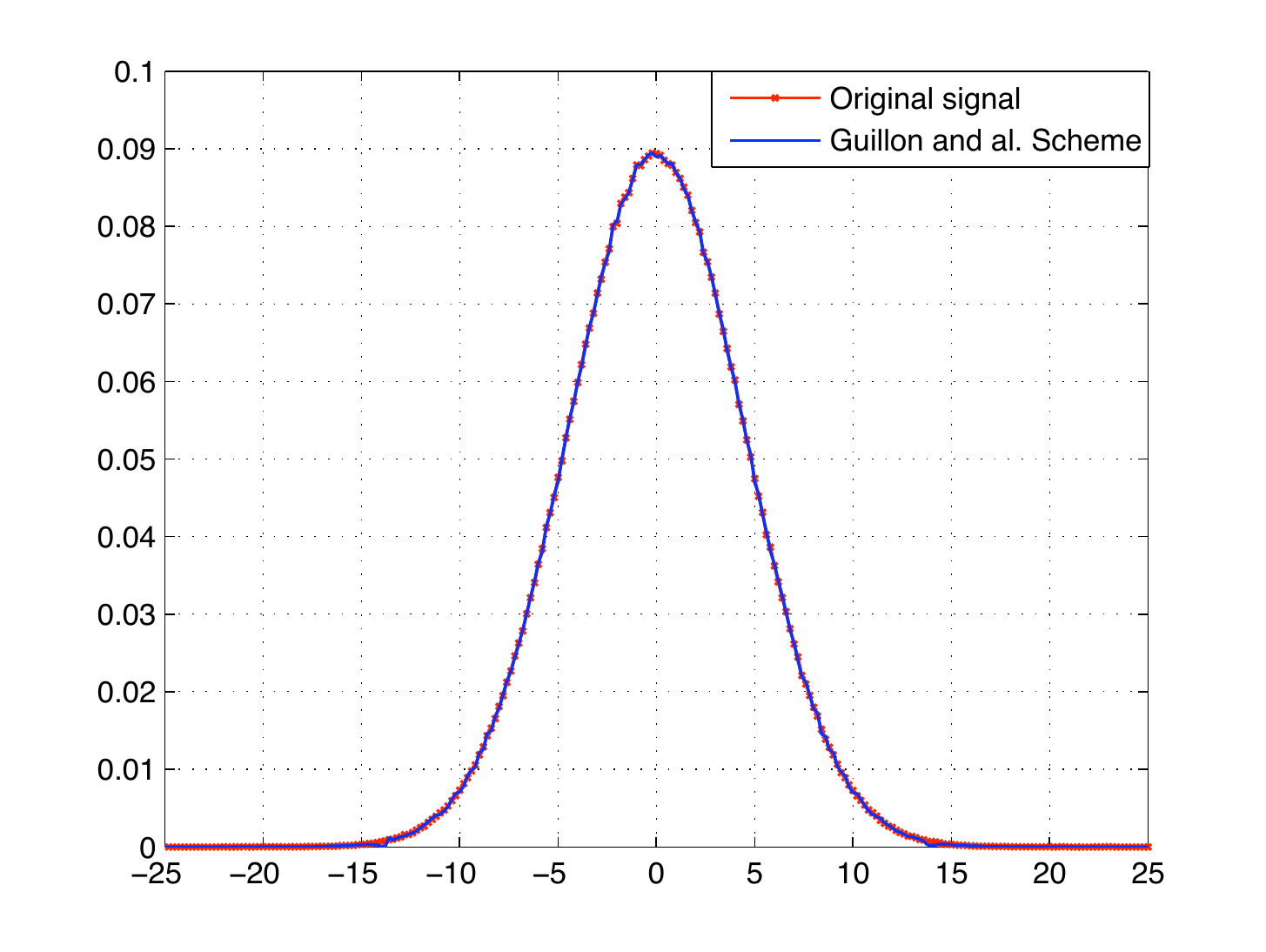}\\
                (a) & (b)
            \end{tabular}
            \caption{(a) Probability density functions
            of the host and marked signal using SCS for document to watermark ratio equal to 13~dB with $\alpha=0.3$ and (b)
                    the probability density function of
                    stego-signal with Guillon \emph{et al.} scheme
                    and the original cover-signal.}
            \label{fig_ddp_SCS}
        \end{center}
    \end{figure}

For convenience, $\textbf{u}^\star[i]$, $\textbf{m}[i]$ and
$\textbf{x}[i]$ are denoted respectively as $u$, $m$ and $x$ in this
section. If the information bits are equiprobable, then (see
appendix~\ref{appendixA}):
    \begin{equation}
        p_X(x) = \frac{1}{2(1 - \alpha)} \sum_{u, m} 1_{\left[
                u - \frac{(1 - \alpha) \Delta} 2 ,
                u + \frac{(1 - \alpha) \Delta} 2
            \right]} p_S \left(
            \frac{x - \alpha u}{1 - \alpha}
        \right) \textrm ,
       \label{eq1}
    \end{equation}
where $1_{\left[ . \right]}$ represents an unit window function. In
this case, the distance between the reconstruction points of the two
quantizers is equal to $\Delta / 2$, and then any window function
recover the nearest ones if $(1 - \alpha) \Delta / 2 > \Delta / 4$
(which is equivalent to $\alpha < 1/2$); and for $\alpha > 1/2$ the
window functions are separated. This explains the aliasing in the
p.d.f.~--for $\alpha=0.3$~-- of the host signal in Fig.~\ref{fig_ddp_SCS}(a).

For $\alpha = 1/2$, there are no holes and no aliasing but we obtain
a continuous p.d.f. only if $p_X(u / 2) = p_X(u / 2 + \Delta / 4)$.
The last equality is satisfied only if the p.d.f. is uniform.\\

The observed discontinuities lead to a statistical detectable
embedding. In the next part, we propose to study an improved scheme
based on SCS.

            \subsection{Improvement of SCS: Guillon \emph{et al.} scheme}

\begin{figure}
         \begin{center}
                \includegraphics[width={.5\textwidth} ,height=4cm]{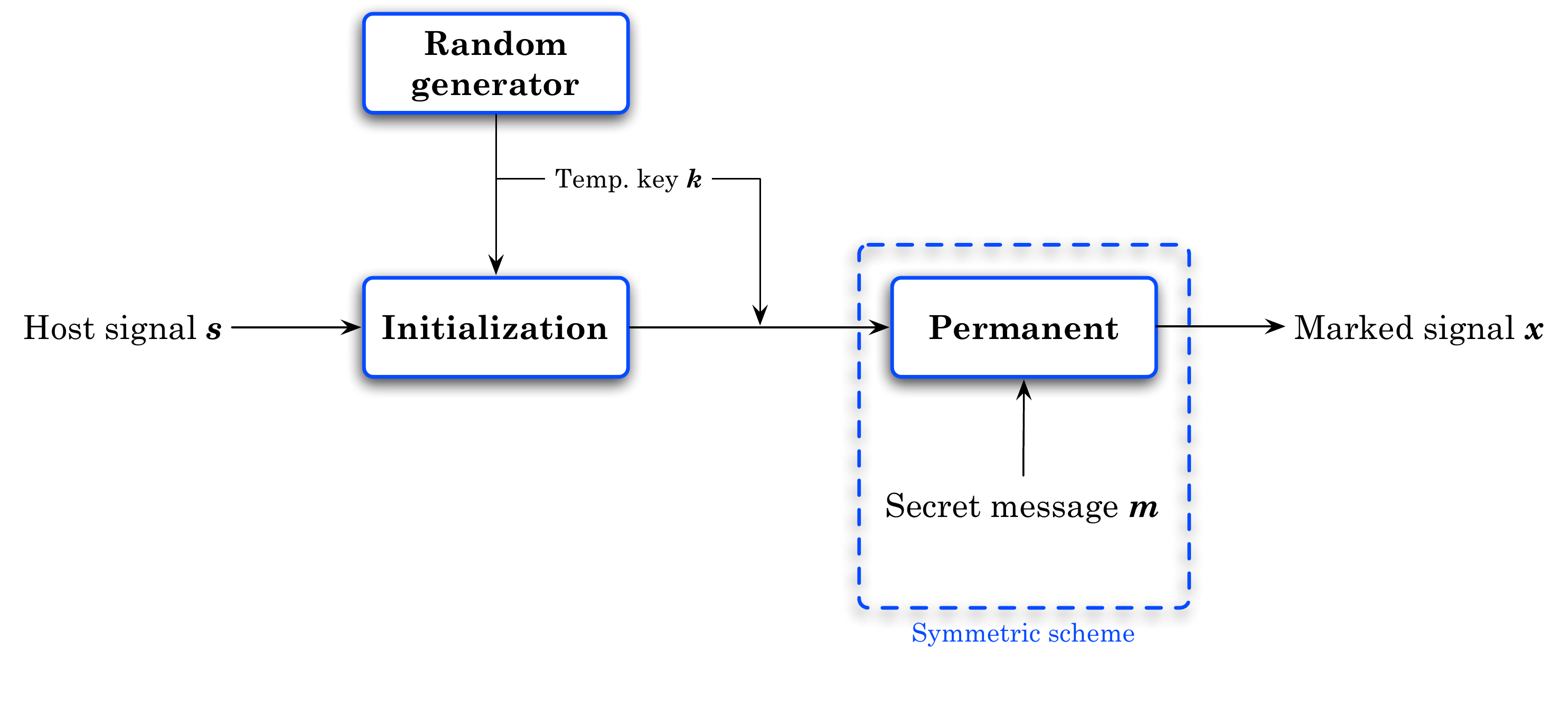}
            \caption{Asymmetric steganography
                    scheme: the permanent phase is initialized
                    with a temporary private key $\textbf{k}$. }
            \label{fig_guillon}
        \end{center}
    \end{figure}

By learning from Anderson and Petitcolas's work~\cite{anderson98},
Guillon \emph{et al.}~\cite{guillon02} proposed a practical scheme
of steganography with public key using asymmetric cryptography and
SCS. Fig.~\ref{fig_guillon} summarizes the two phases of this
scheme. In the initialization phase, a private key $\textbf{k}$ is
generated with a pseudo-random generator and is encrypted with an
asymmetric cypher algorithm. The key $ C(\textbf{k},
\textbf{k}_{\scriptsize \textrm{pub}})$ --~where $
\textbf{k}_{\scriptsize \textrm{pub}}$ is a public key known by all
users~-- is embedded on the cover-signal. The permanent phase uses
the transmitted key $\textbf{k}$ and SCS to embed and transmit the
message $\textbf{m}$.

In the permanent phase, the statistical undetectability is mainly
assured by the private key, since it leads to a non distorted
p.d.f. However, the initialization phase requires the transmission of
public information without distorting the stego-signal. Guillon
\emph{et al.} proposed to use SCS with $\alpha = 1/2$ in order to
hide an invisible (statistically and perceptually) message, but it
is only valid for a cover-signal with uniform p.d.f.; they then
proposed to use a compressor before embedding in order to equalize
the p.d.f. of cover-content. The embedded message will be
statistically invisible, as shown in Fig.~\ref{fig_ddp_SCS}(b).
Unfortunately, the resulting stego-system is less flexible, because
the encoding and decoding steps highly depend on the statistics of
cover-content. It has been recently shown~\cite{label_ref_Gaetan}
that the artifacts in the stego-signal are due to the use a regular
partitioning codebook. In the next section, we propose to use a
structured codebook by the way of TCQ.

        \section{Analysis of the trellis-coded quantization}

The approach proposed here concerns the use of a trellis-based
quantization, for a pseudo-random partitioning of the codebooks, in
order to avoid the artifacts introduced in the p.d.f. of
stego-content by regular partitioning (as observed in the previous
stego-system).

              \subsection{Principles}

Let us consider a trellis defined by a transition function: $
\mathcal{E} \times \{ 0, 1 \}  \longrightarrow  \mathcal{E}$,
$\textrm{tr} : \left( \textbf{e}[i], \textbf{m}[i] \right)
\longmapsto \textbf{e}[i+1] $, with $\mathcal{E} = \{ 0, 1, \ldots ,
2^{r-1} \}$ groups of possible states, where $r$ is an integer such
as $r > 1$, and $i$ is the index of current transition. Contrary to
the SCS, the dithering $\textbf{d}$ will not be random but will
become a function of the current state and of the embedded symbol:
\begin{eqnarray}
    \mathcal{E} \times \{ 0, 1 \} &\longrightarrow&
      [- \Delta / 2 , + \Delta / 2] \textrm , \nonumber \\
    f : \left(
                 \textbf{e}[i], \textbf{m}[i]
            \right)  &\longmapsto&  \textbf{d}[i] \textrm .
\end{eqnarray}
In this stego-system, the codebooks are defined by
    $$
     \mathcal{U}_{\textbf{m}}[i] = \left\{ n \Delta + f(\textbf{e}[i], \textbf{m}[i]) \textrm{,~} n \in  \mathcal{Z} \right\} \textrm ,
    $$
and the closest codeword $\textbf{u}^{\star} \in
\mathcal{U}_{\textbf{m}}$ to $\textbf{s}[i]$ is calculated using a
Viterbi algorithm~\cite{viterbi}, with a high \emph{a priori} in
order to be sure that the obtained codeword belongs to
$\mathcal{U}_{\textbf{m}}$:
        \begin{equation}
             \textbf{u}^\star = \arg \min_{\textbf{u} \in \mathcal{U}_{\textbf{m}}} \sum_{j = 1}^G \left(
                 \textbf{s}[j] - \textbf{u}[j]
            \right)^2 \textrm .
            \label{eq:ustar}
        \end{equation}
The stego-signal is given by:
    \begin{equation}
     \textbf{x} = \textbf{s} + \alpha \left(
        \textbf{u}^{\star} - \textbf{s}
    \right) \textrm ,
    \end{equation}
where $\textbf{s}$ is the cover-signal and $\alpha$ represents the
Costa's parameter.\newline To extract the embedded message, we have
to apply the Viterbi algorithm in order to retrieve the path which
corresponds to the stego-signal.

            \subsection{Statistical analysis of TCQ}

\begin{figure}
        \begin{center}
            \begin{tabular}{cc}
                \includegraphics[width=0.22\textwidth ,height=3.5cm]{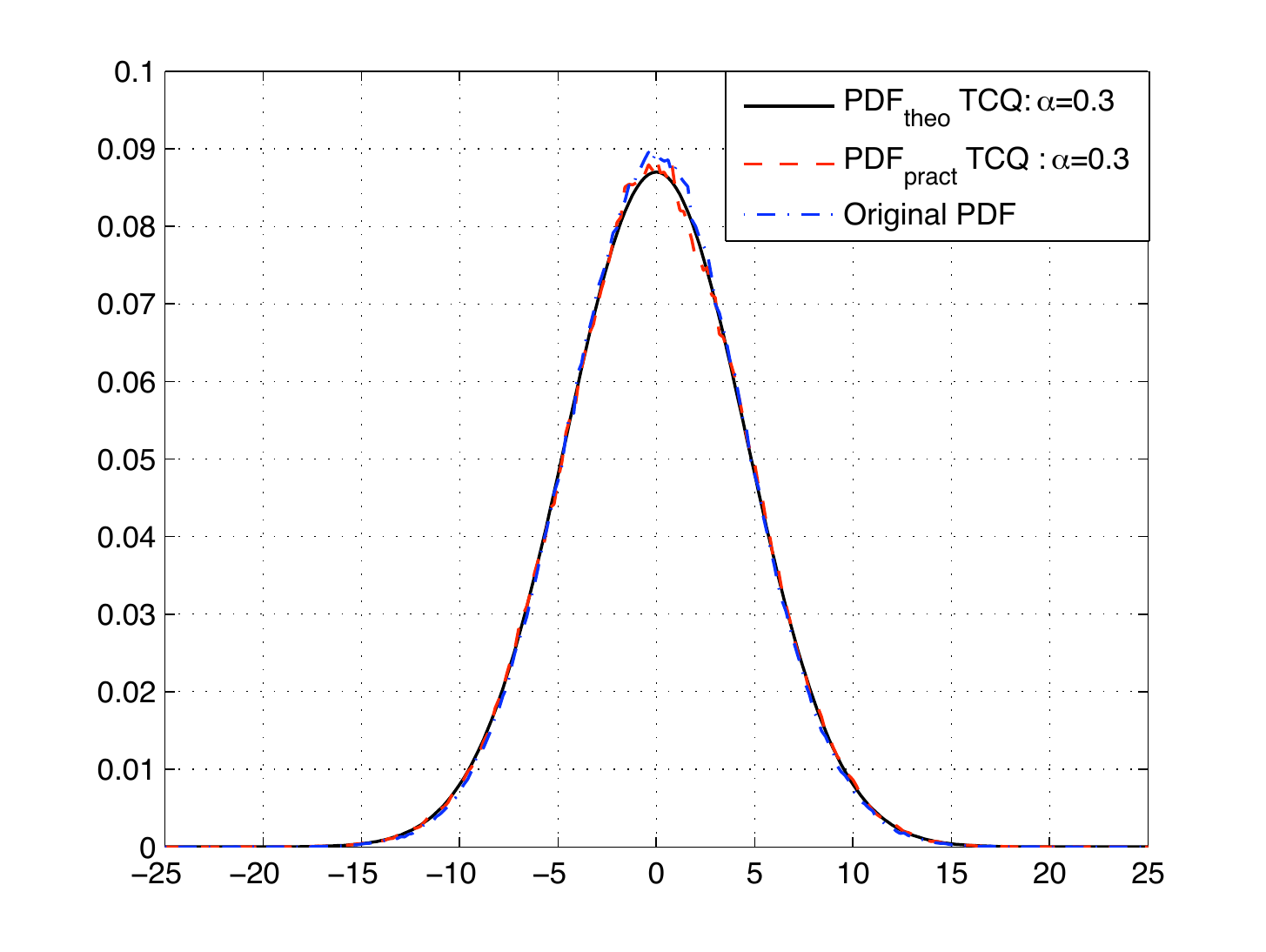} &
                \includegraphics[width=0.22\textwidth ,height=3.5cm]{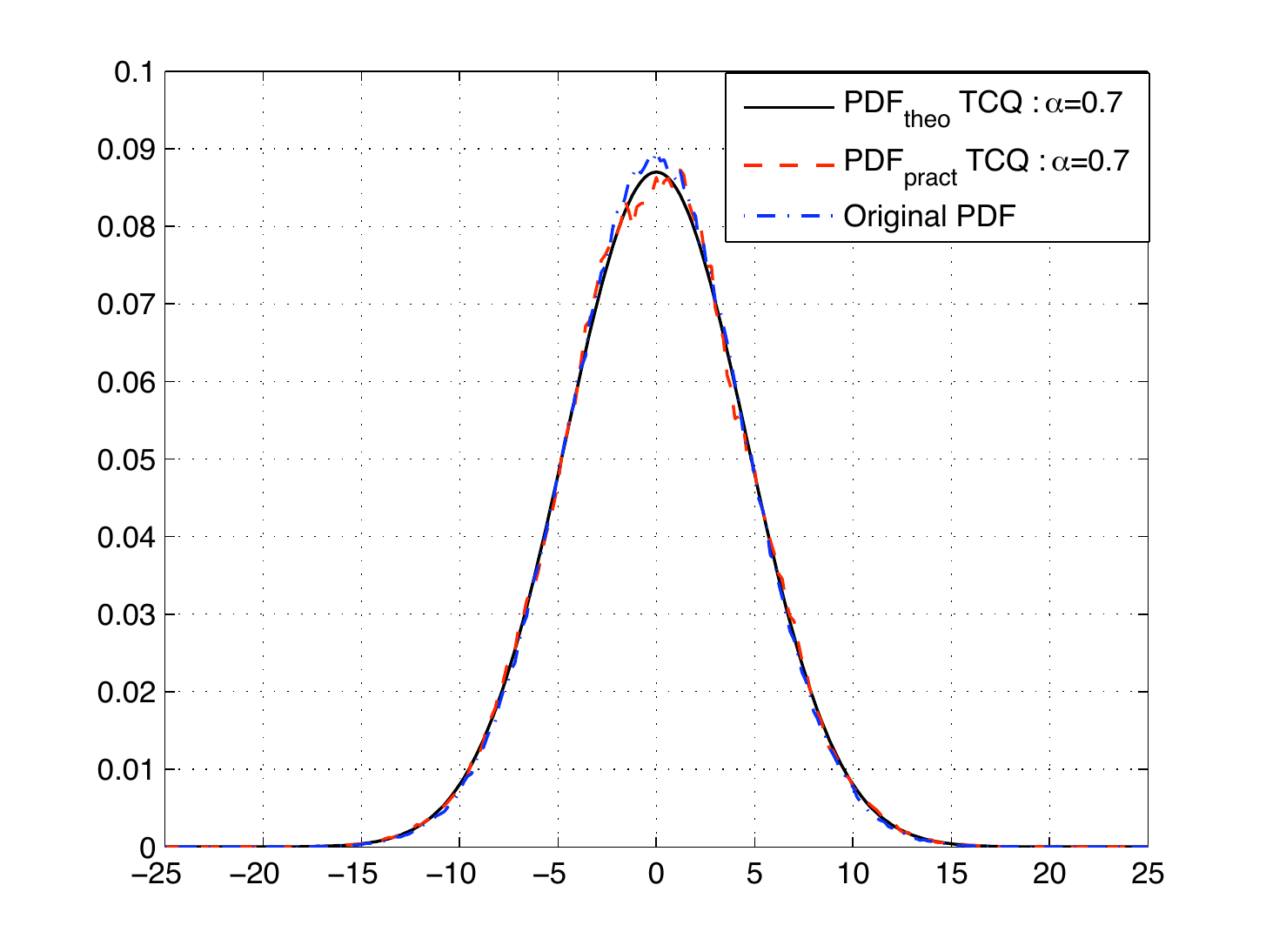}\\
                (a) & (b)
            \end{tabular}
            \caption{Probability density functions
            of the cover and stego-signal for document to watermark raio equal to 13~dB by using TCQ with different
            value of $\alpha$: (a) $\alpha=0.3$
            and (b) $\alpha=0.7$.}
            \label{fig_TCQ_ddp}
        \end{center}
    \end{figure}

In order to theoretically justify the use of the TCQ to get
statistical invisibility, we have calculated the p.d.f. (see
appendix~\ref{appendixB}). We obtain:
    \begin{equation}
        p_X(x)
        = \frac{1}{\sigma_{W}\sqrt{12}}\int_{x-\sigma_{W}\sqrt{3}}^{x+\sigma_{W}\sqrt{3}}p_{S}(z) \textrm{~d} z \textrm ,
        \label{eq:tcqlss}
    \end{equation}
where $\sigma_W$ is the standard deviation of the embedded signal.
Then $p_X$ is the mean p.d.f. for the cover signal in the interval
centered on $x$ and a width $\sigma_W \sqrt 3$. We have implemented
Eqn.~(\ref{eq:tcqlss}) for a signal with Gaussian p.d.f. and we
obtained the results presented on Fig.~\ref{fig_TCQ_ddp}(a) and
\ref{fig_TCQ_ddp}(b). We can notice the good match between the
p.d.f. obtained with the TCQ algorithm (experimental), the
theoretical versions and the original ones for the same high
embedding power.

However, Fig.~\ref{fig_cap_dDKL}(a) shows that the capacity of TCQ
is not as good as that of SCS. Then, we can use the TCQ only in the
initialization phase~-- of the previous scheme
(Fig.~\ref{fig_guillon})~--, because this phase requires just a
limited payload.

            \section{Analysis of the spread transform scalar Costa scheme}

We propose to use the ST system which allows any stego-system to
increase its Watermark-to-Noise Ratio (WNR)~\cite{eggers03} and improve the resistance against active
warden (who performs an AWGN attack in order to remove the
stego-message).

                \subsection{Spread transform}

Chen and Wornel~\cite{label_ref_ChenWaornel} introduced a general
approach for robust watermarking applications. It allows to spread
the embedded message on several cover samples. They proposed to hide
the message in a transformed domain~\cite{eggers03}. In sequel, the spreading parameter is modeled by a realizations set
of random variables with uniform p.d.f. To extract the hidden
message, an inverse transformation is applied to a resulted signal.

In~\cite{eggers03}, authors studied especially the robustness of
this system to applied it to the robust watermarking. In this work,
we study the steganographic performance of the spread transform
system in active warden context. We note that, before transmitted
the information, the spread transform makes an inverse
transformation where the embedded signal strength is divided by the
spreading factor $\tau$, then $\textrm{DWR} = \textrm{DWR}_{\tau} + 10 \log_{10} \tau$,
such as $\textrm{DWR}$ is the Document-to-Watermark Ratio and
$\textrm{DWR}_{\tau}$ is the Document-to-transformed Watermark
Ratio. Thus, spread transform improves the perceptual invisibility
of any hiding system.

  \begin{figure}
         \begin{center}
            \begin{tabular}{c}
                \includegraphics[width=0.35\textwidth]{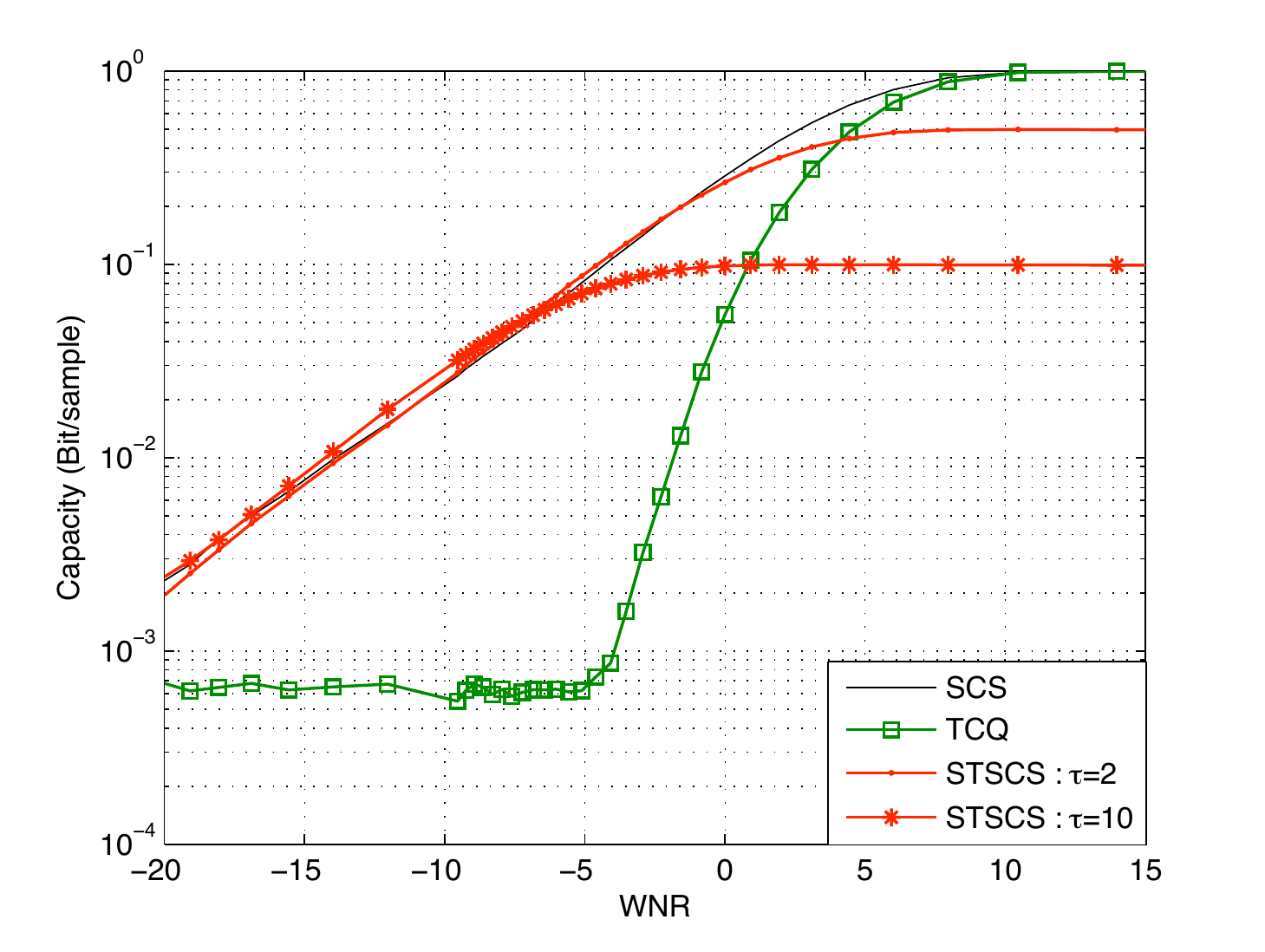}
\\
(a)\\
                \includegraphics[width=0.35\textwidth]{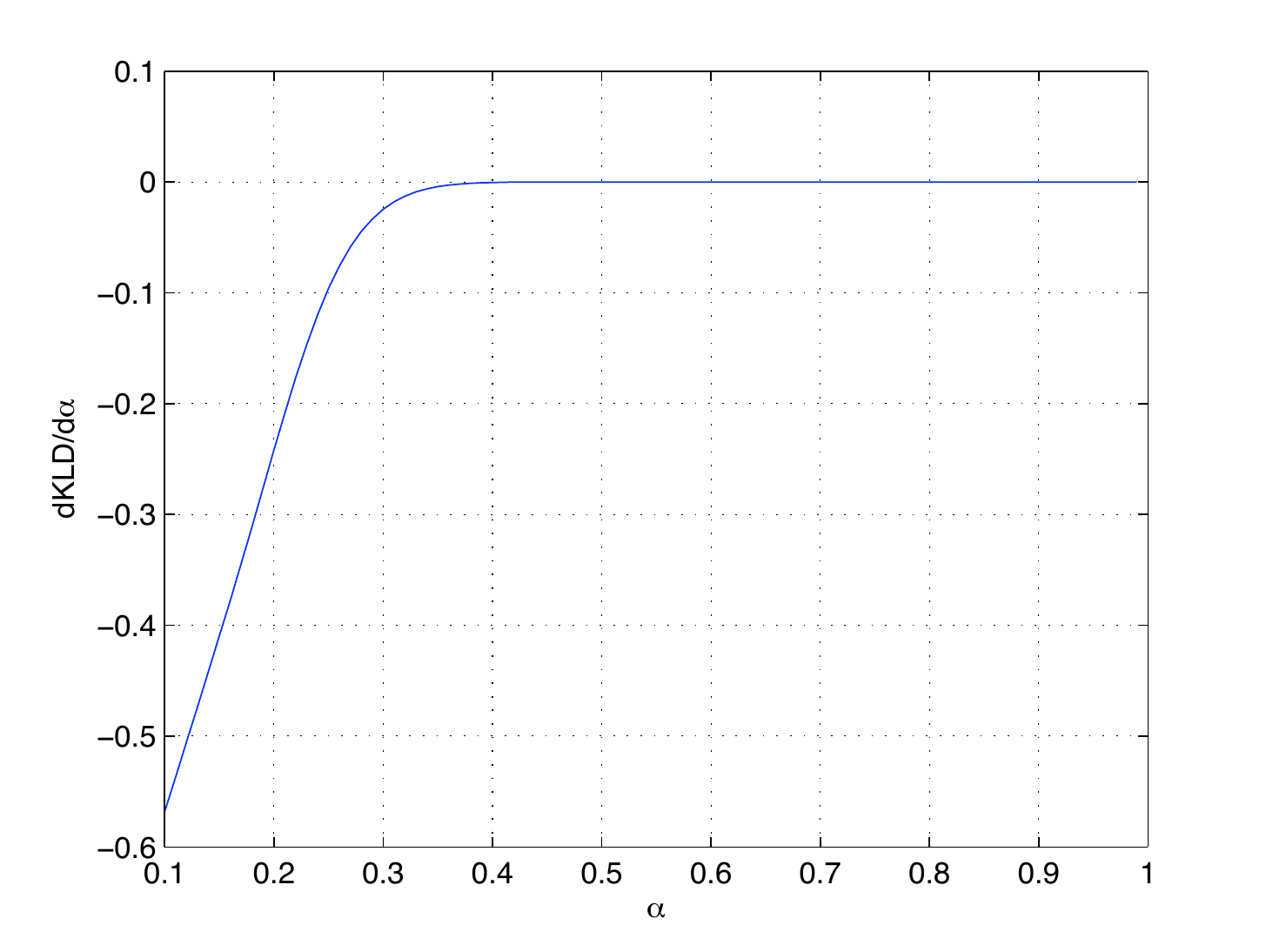} \\
                (b)
            \end{tabular}
            \caption{(a) The capacity of stego-systems
                SCS, TCQ and ST-SCS as function of watermark to noise ratio and (b)
                the differentiation of the function $\textrm{kld}(\alpha)$ with respect to
                the parameter $\alpha$, in the case of ST-SCS stego-system with $\tau = 2$.}
            \label{fig_cap_dDKL}
        \end{center}
    \end{figure}

            \subsection{Statistical analysis of ST-SCS}

\begin{figure}
        \begin{center}
            \begin{tabular}{ccc}
                \includegraphics[width=0.22\textwidth ,height=3.5cm]{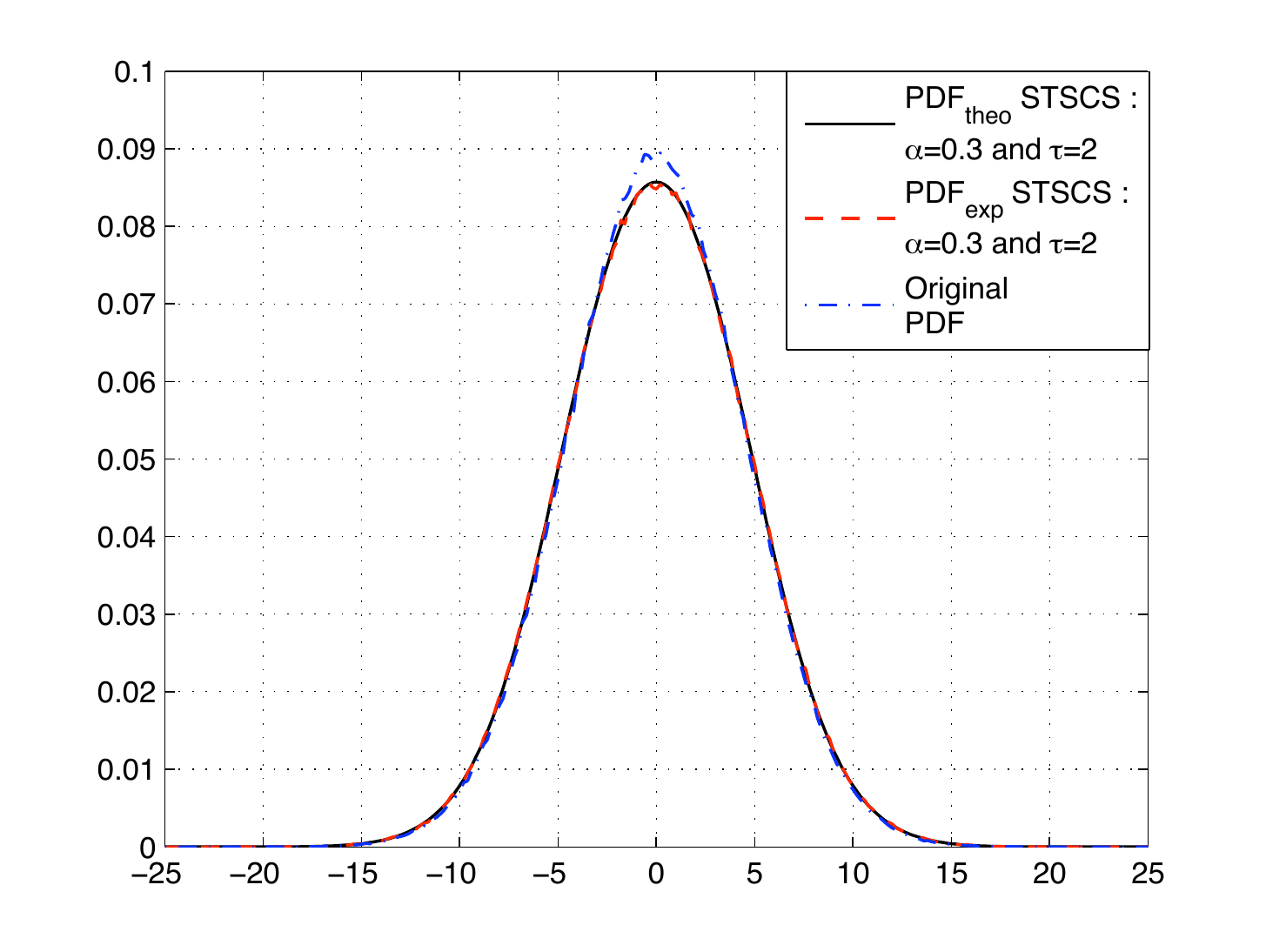} &
                \includegraphics[width=0.22\textwidth ,height=3.5cm]{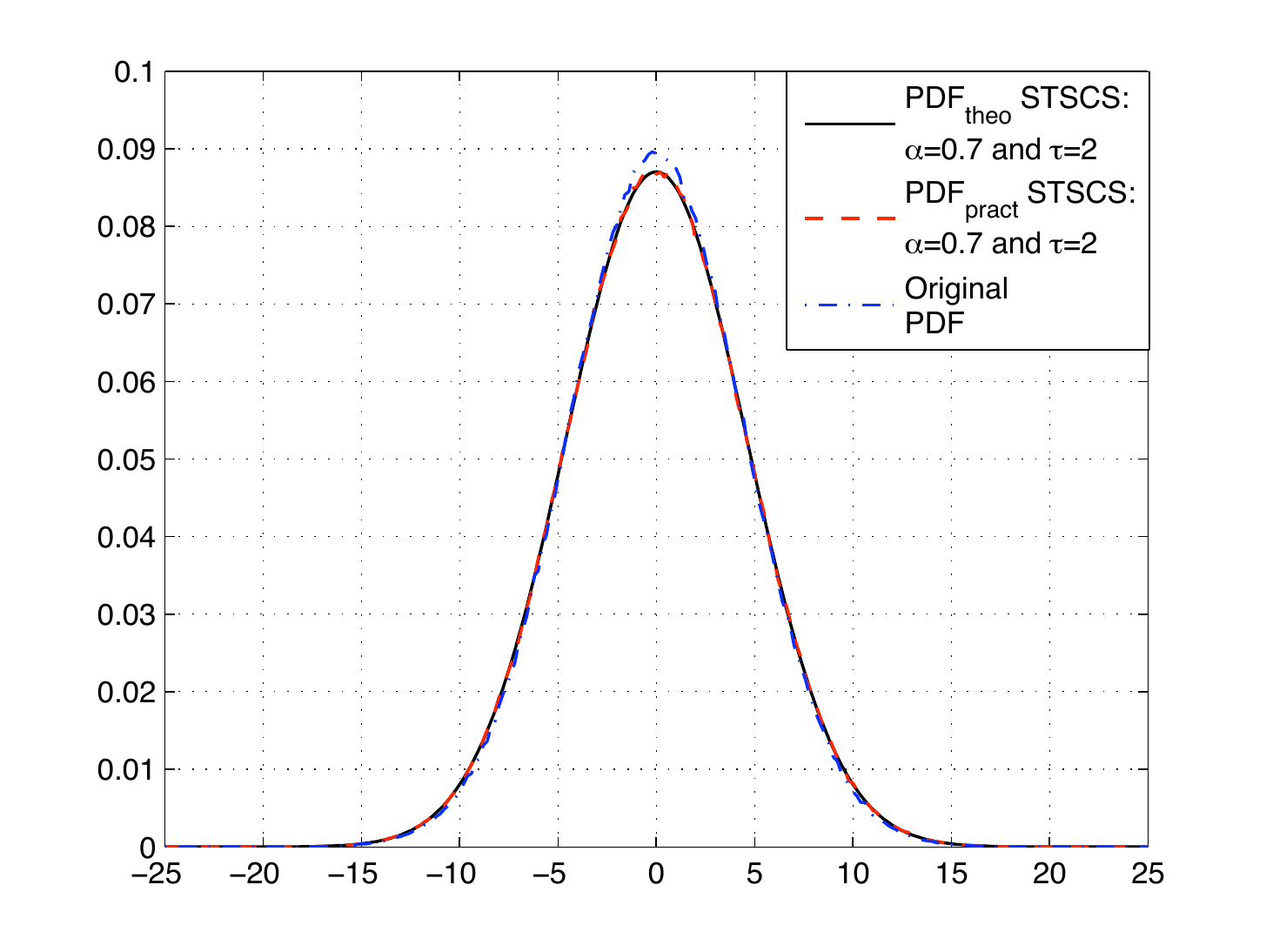} \\
                (a) & (b) \\
                \includegraphics[width=0.22\textwidth ,height=3.5cm]{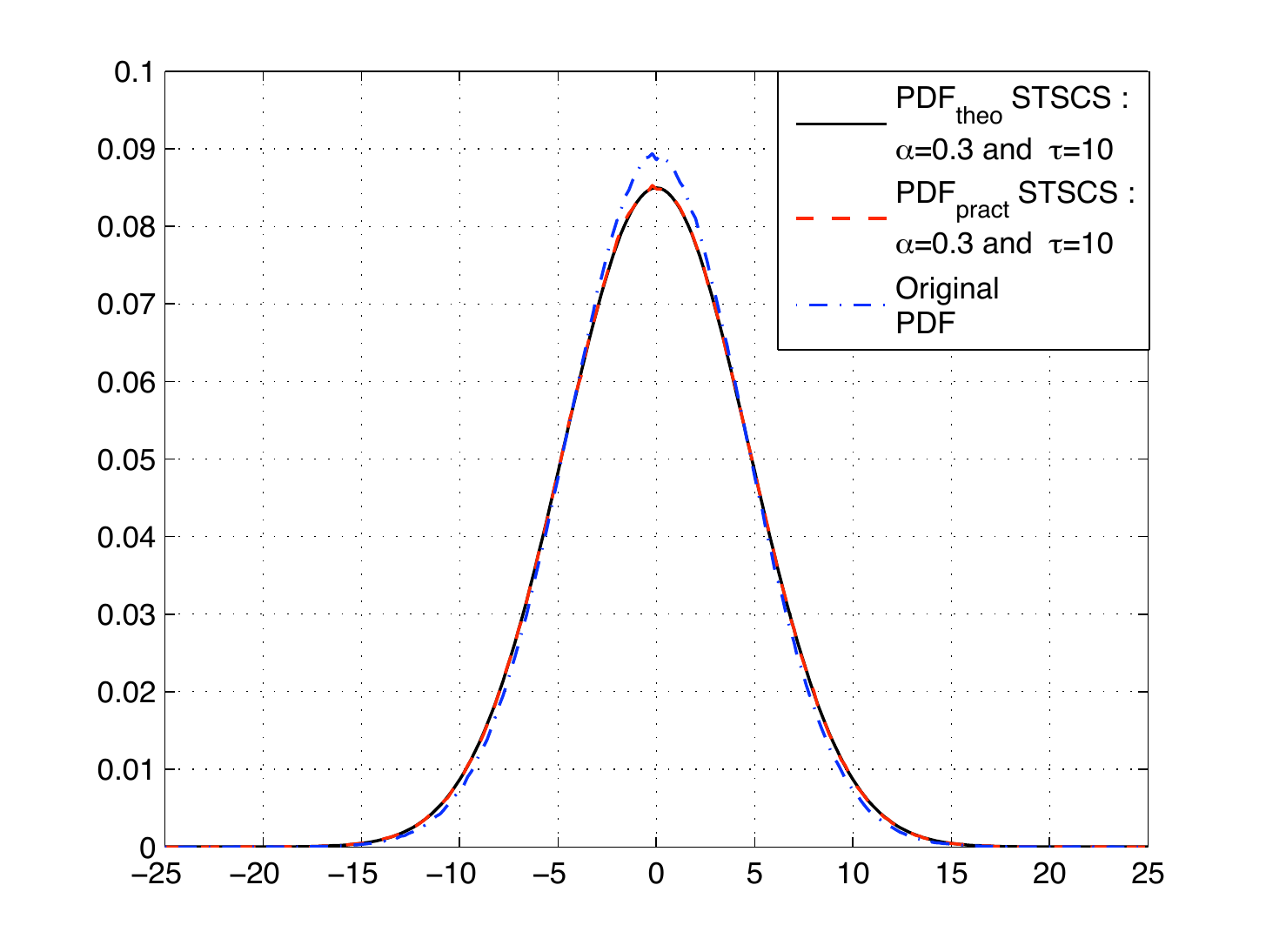} &
                \includegraphics[width=0.22\textwidth ,height=3.5cm]{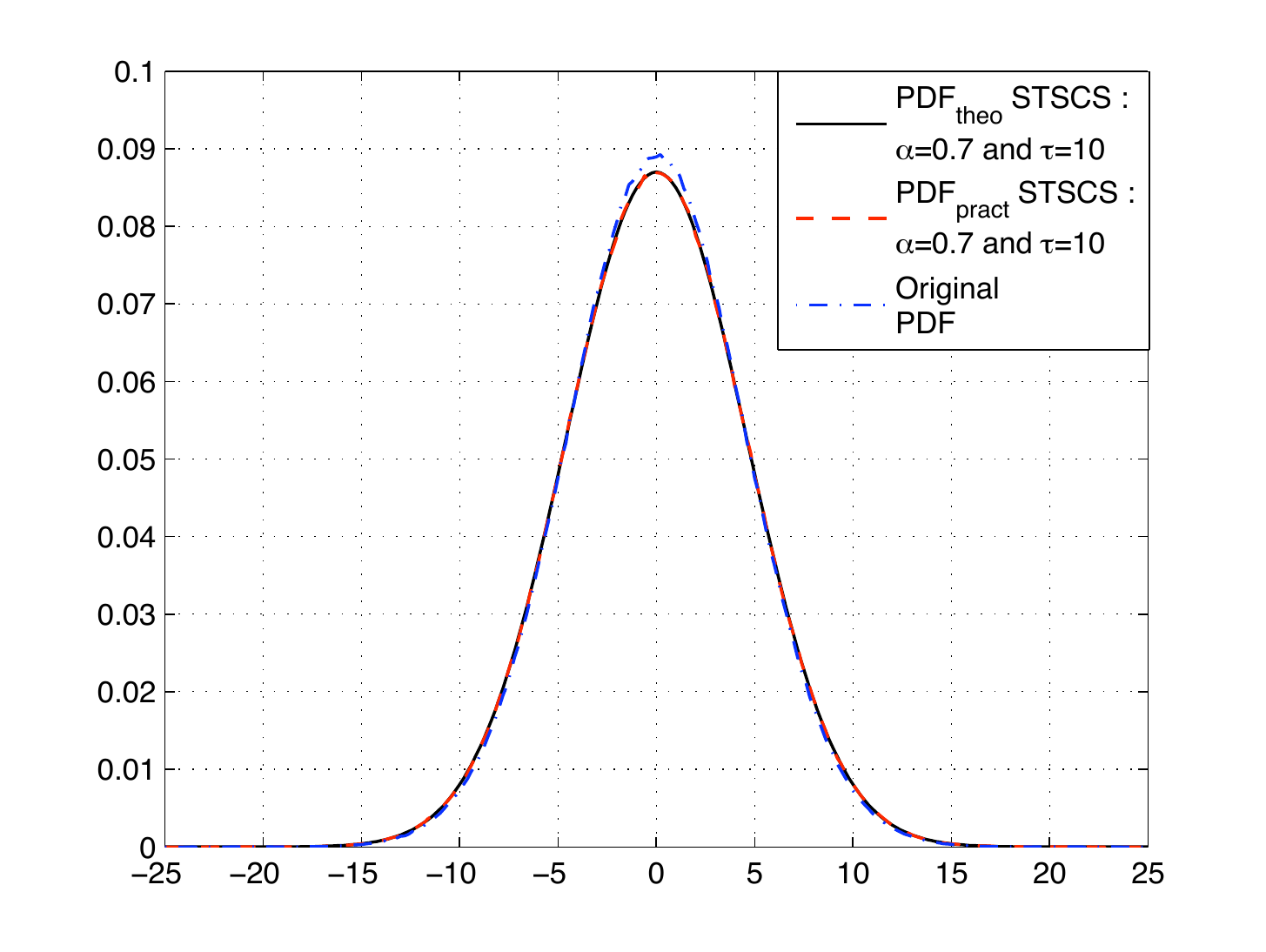} \\
                (c) & (d)
            \end{tabular}
            \caption{Probability density functions
            of the cover and stego-signal by using ST-SCS
            for $\tau = 2$ and document to watermark ratio equal to 13~dB with different value of $\alpha$: (a)
            $\alpha = 0.3$ and (b) $\alpha = 0.7$;
            for $\tau = 10$ with (c)
            $\alpha = 0.3$ and (d) $\alpha=0.7$.}
            \label{fig:ddp}
        \end{center}
    \end{figure}

In sequel, we focus only on the combination of the spread transform
with the SCS-based stego-system in active warden context. In order
to evaluate the statistical undetectability of the stego-system, we
develop a theoretical formulation of ST-SCS stego-signal density
(see appendix~\ref{appendixC}):
\begin{eqnarray}
    && p_{X}(x)  =  \frac \tau {4 (\tau - \alpha)}  \nonumber\\ && \sum_{u, m, t} \int_y \delta \left(
        u - Q_{\Delta} \left(
            \frac \tau {\tau - \alpha} \left(
                x + \alpha y t - \alpha u t
                \right) t + y
            \right)
        \right) \nonumber \\
        && \times p_{s} \left(
            \frac \tau {\tau - \alpha} \left(
                x + \alpha y
                t - \alpha u t
            \right)
        \right) p_{Y}(y) \textrm{~d} y \textrm . \nonumber \label{eq_st}
\end{eqnarray}
In Fig.~\ref{fig:ddp}, the experimental p.d.f. of the stego-signal validates the theoretic model given by Eqn.~\ref{eq_st}, because we can see that the theoretic p.d.f. follows the experimental one. \\

\begin{figure}
         \begin{center}
            \begin{tabular}{cc}
                \includegraphics[width=0.35\textwidth]{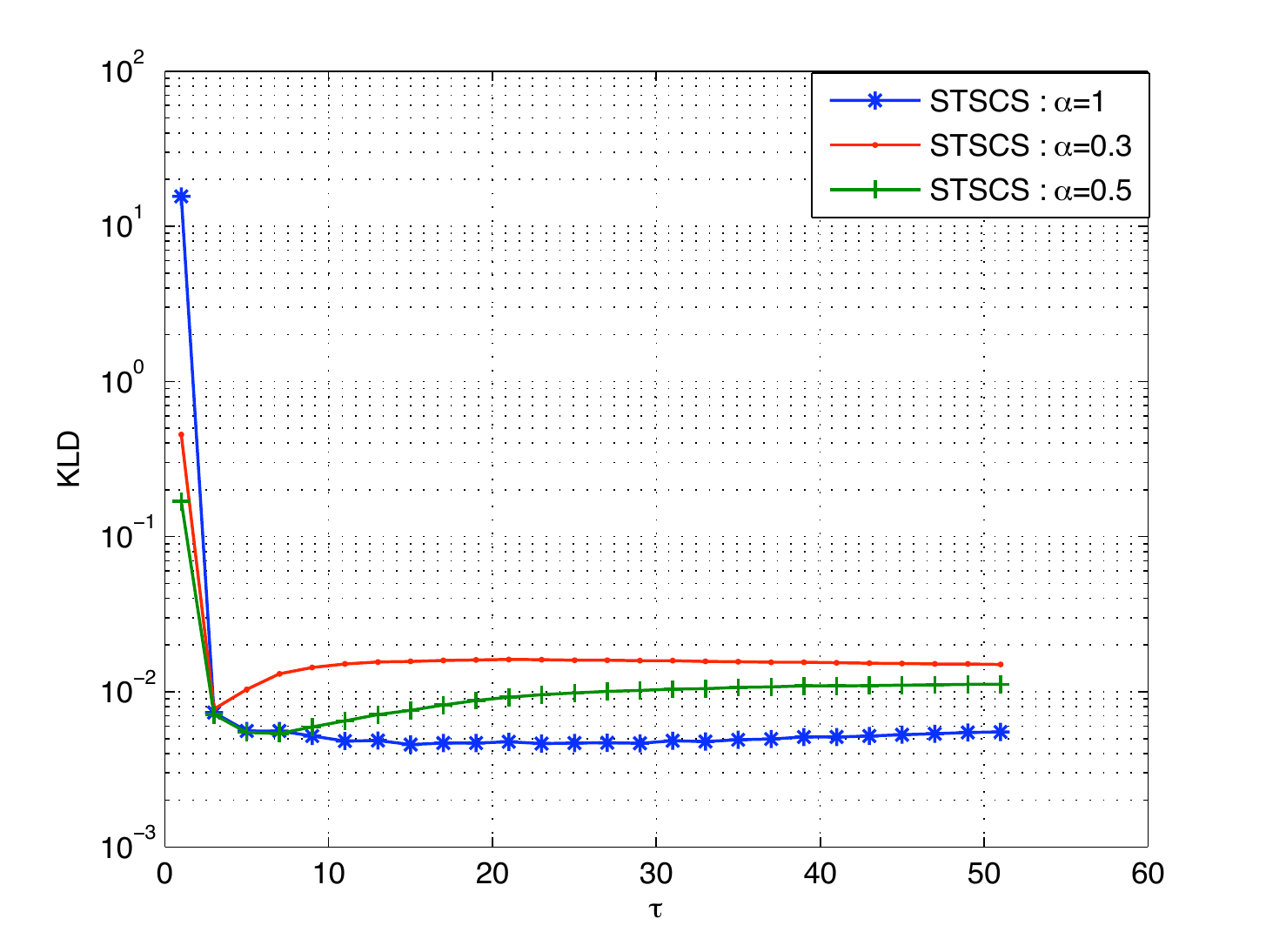}\\ (a) \\
                \includegraphics[width=0.35\textwidth]{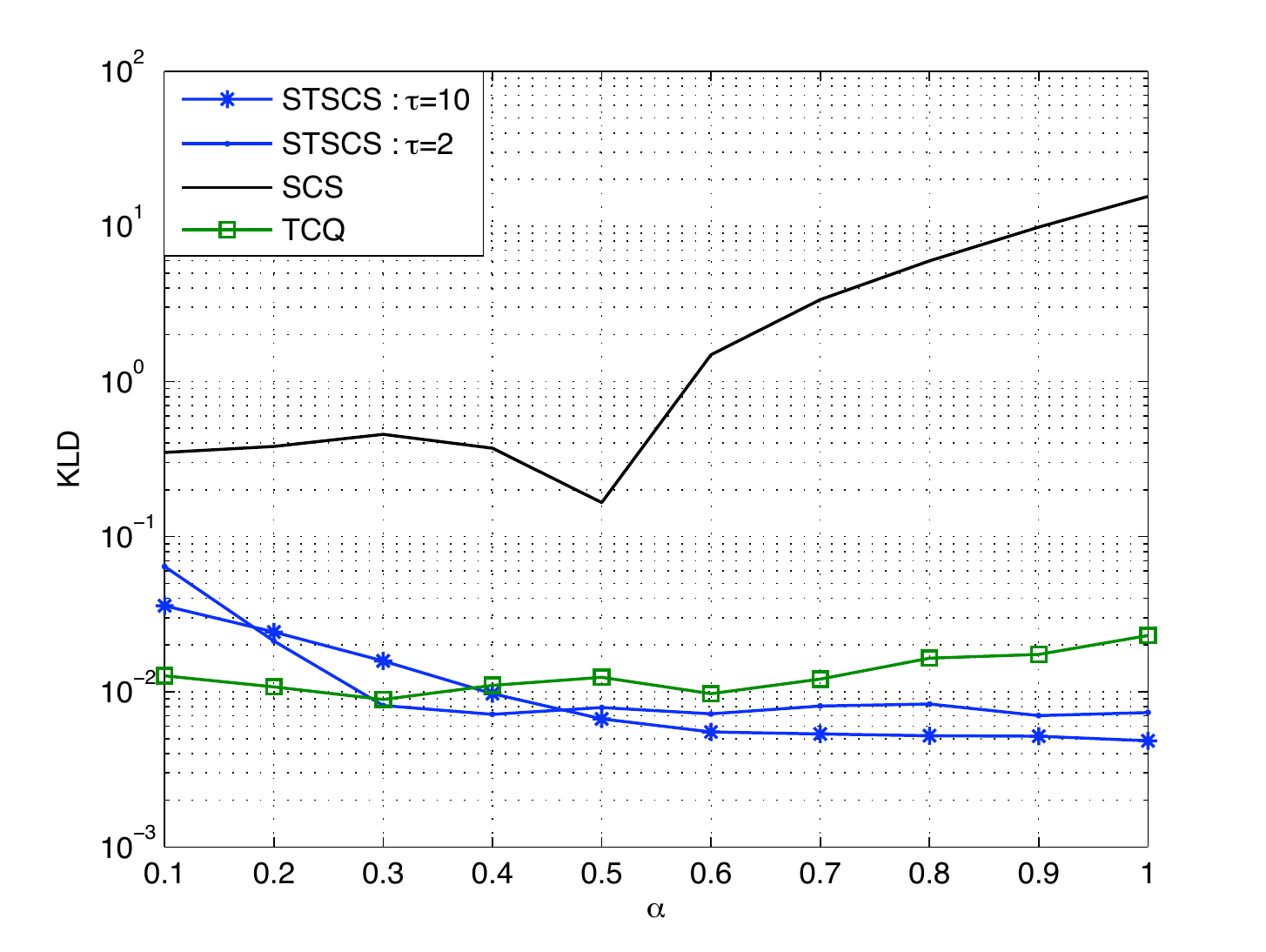} \\
                 (b)
            \end{tabular}
            \caption{(a) The Kullback-Leibler distance of ST-SCS as function of $\tau$
                for different value of $\alpha$ and (b) the Kullback-Leibler distance as function of $\alpha$
                for different value of $\tau$.}
            \label{fig_ST-SCS}
        \end{center}
    \end{figure}

If we replace $t$ with its two possible realizations, i.e. $\pm 1 / \sqrt \tau$, and we take $\tau \rightarrow \infty$ with finite $\sigma_{s}^2$ (the variance of cover-signal $s$) then:
\begin{eqnarray}
    p_{X}(x) & = & \frac 1 4
        \sum_{u, m} \int_{y} \delta \left(
            u - Q_{\Delta} \left(
                y
            \right)
        \right) p_{S}(x) p_{Y}(y) \textrm{~d} y \nonumber \\
        && + \frac 1 4 \sum_{u, m} \int_{y} \delta \left(
            u - Q_{\Delta}(y)
        \right) p_{S}(x) p_{Y}(y) \textrm{~d} y
    \textrm . \nonumber 
\end{eqnarray}
So the stego-signal $\textbf{x}$ has the same density as the
cover-signal~-- in this case the two p.d.f. are both Gaussian. However,
Fig.~\ref{fig_cap_dDKL}(b) shows that the differentiation of the
KLD by respect to $\alpha$ is always negative and converges
speedily to zero even for $\tau = 2$, then the KLD takes~--
theoretically~-- its minimal value for the majority values of the
parameter $\alpha$ and for any value of the spreading factor $\tau$.
In addition, experiences show that the
 stego-signal has the same p.d.f. than the cover-signal even
for a small value of spreading factor $\tau$ (see
Fig.~\ref{fig:ddp}). We can see on Fig.~\ref{fig_cap_DKL_DWR}(a),
Fig.~\ref{fig_ST-SCS}(a) and Fig.~\ref{fig_ST-SCS}(b) that ST-SCS
has the same level of the statistical undetectability as TCQ
stego-system, but better than the undectability level of the SCS.

            \subsection{Performance of ST-SCS}

\begin{figure}
         \begin{center}
            \begin{tabular}{cc}
                \includegraphics[width=0.35\textwidth, height=5cm]{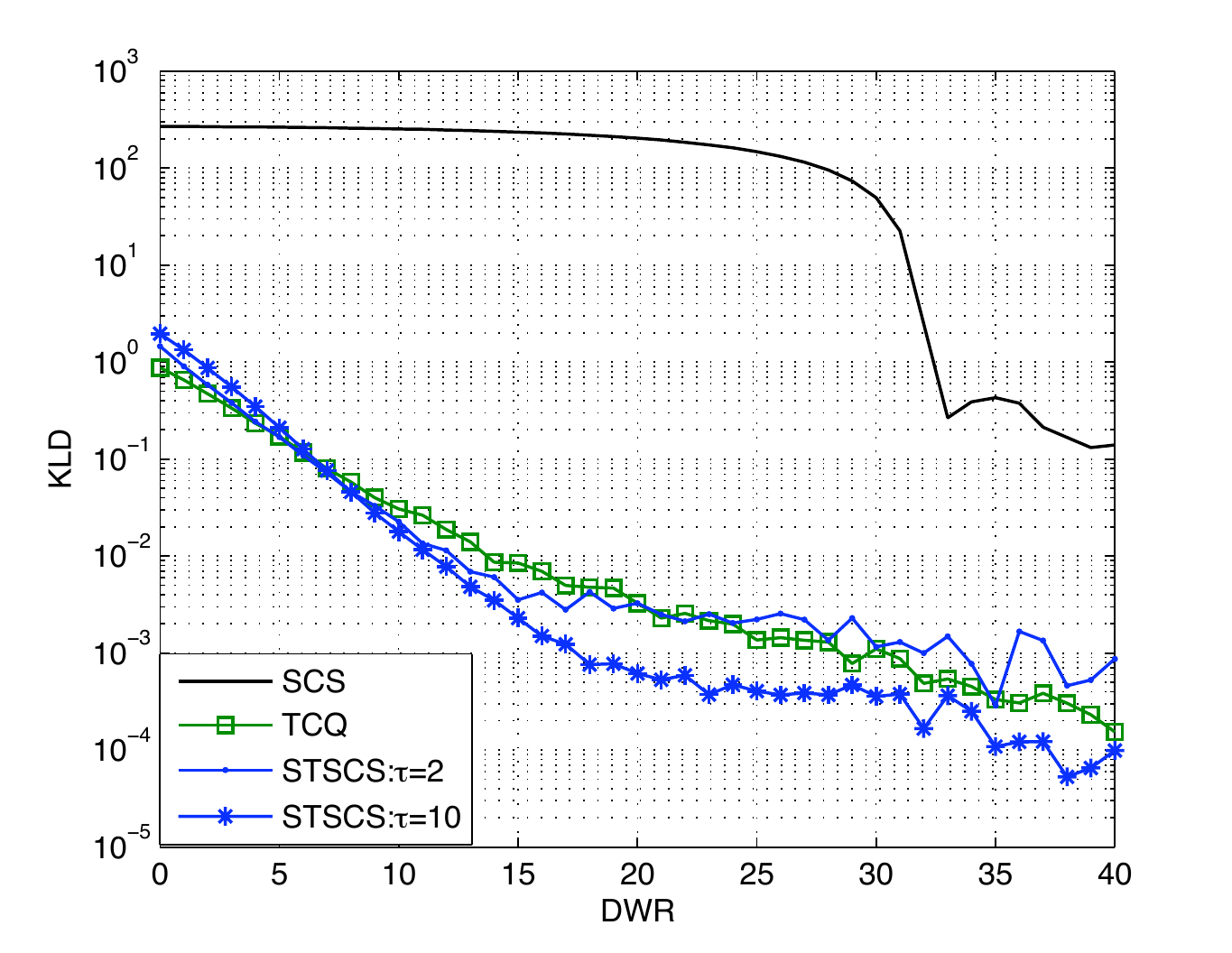}
\\
(a)\\
                \includegraphics[width=0.35\textwidth, height=5cm]{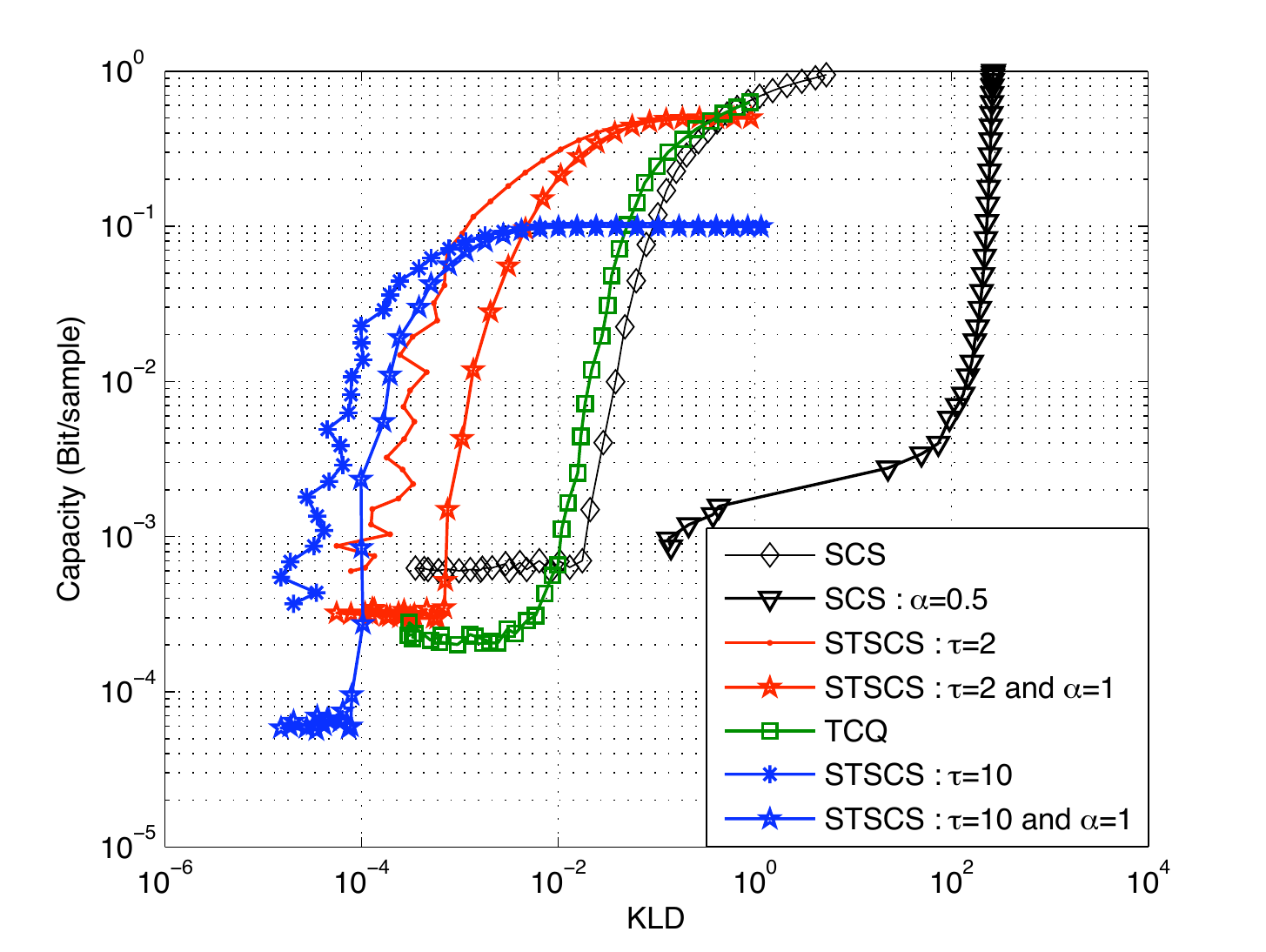} \\
                (b)
            \end{tabular}
            \caption{(a) The Kullback-Leibler distance for SCS, TCQ
                and ST-SCS stego-systems with Gaussian images as function of DWR;
                (b) capacity vs. Kullback-Leibler distance for SCS, TCQ
                and ST-SCS stego-systems with Gaussian images such that WNR $\in [-20, 12]$~dB and document-to-watermark ratio $\in [0, 40]$~dB.}
            \label{fig_cap_DKL_DWR}
        \end{center}
    \end{figure}

Fig.~\ref{fig_cap_dDKL}(a) shows that for strength warden attack
(low WNR), the capacity of ST-SCS is better than the one of TCQ.
In the contrary, for high WNR values, the capacity of the TCQ is
better. As a result, it is very difficult to have a system which
permits a good invisibility and in the same time a good capacity;
then the compromise between these two characteristics becomes
important. Fig.~\ref{fig_cap_DKL_DWR}(b) shows that the compromise
of ST-SCS is the best in comparison to the SCS and the TCQ
stego-systems in active warden context.

\begin{figure}
        \begin{center}
                \includegraphics[width=0.35\textwidth]{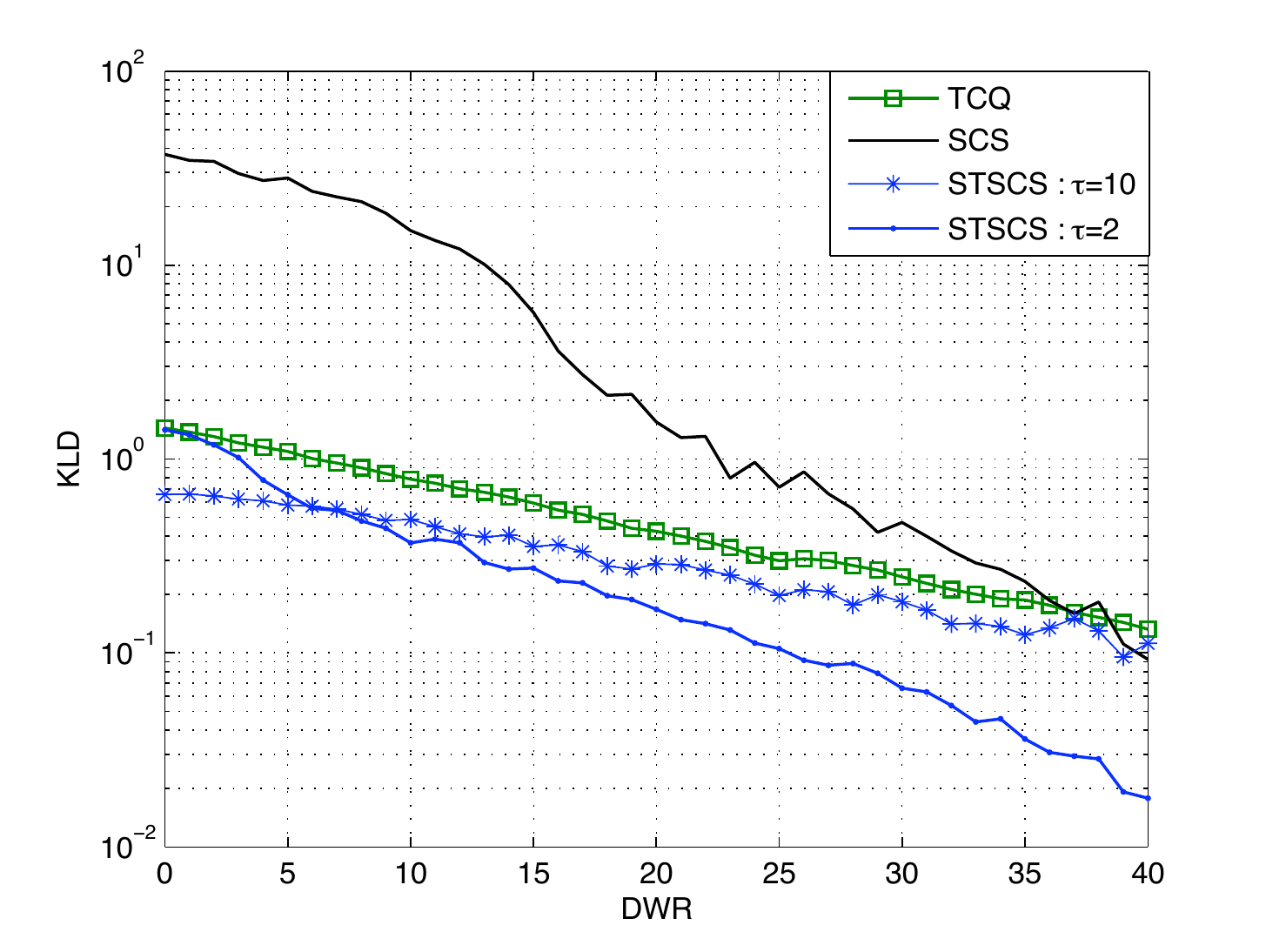}
                \caption{KLD vs. document to watermark ratio with 100 real images of size $350 \times 350$.}
                \label{DKL_real}
            \end{center}
        \end{figure}

We have applied SCS, TCQ-based scheme and ST-SCS
to $100$ real images with $350 \times 350$ pixels size.
Fig.~\ref{DKL_real} confirms the results obtained for Gaussian
images, where the ST-SCS has the same undetectability level as TCQ
and better than SCS. However, the statistical undetectability
will be the same as SCS in transformed domain if the projection parameter is public.\\

In the case of public key steganography (Fig.~\ref{fig_guillon}), we
can use the TCQ stego-system in the initialization phase, to
transmit the secret key, and the ST-SCS in the permanent phase,
which allows to the best
compromise between statistical undetectability and capacity.\\

        \section*{Conclusion and Perspectives}

In this work, we have compared the steganographic performance of
several informed-based stego-systems in active warden context. For
each system, the experimental results have been used to validate the
theoretical model. For SCS, the stego-signal is regularly
partitioned, thus, many artifacts in the p.d.f. of the stego-signal
are introduced, which is also proved by the developed theoretical
formulations. Due to this observations, we have proposed an analysis
of two another systems. The first one is based on a pseudo-random
partitioning (the TCQ-based system), which allows to obtain a more
common and undetectable public stego-system (the technique does not
depend to the cover-signal distribution). The second one is based on
the combination of SCS with spread transform (the ST-SCS), which
allows a good statistical undetectability and a best compromise
between capacity and undetectability. In future work, we shall study
an improvement of the undetectability with combination of ST and TCQ
when the projection parameter is public. We shall also verify our
theoretical models by an applications on real images.

\section{Acknowledgment}
The authors would like to thank Professor Pierre Duhamel for his
help and collaboration to this paper and the
 ESTIVALE project from ANR (French national agency of research) for funding.

        \section{Appendix}

            \subsection{Demonstration of Eqn.~(\ref{eq1})} \label{appendixA} 

We model the stego-signal by a realizations set of
Gaussian random variables, independent and non stationary:
$\mathcal{X} = \{ X[1], \ldots , X[G] \}$. It is given by the
following equation (in sequel, we do not use the index of the
variable for ease of presentation):
\begin{equation}
    X = ( 1 - \alpha) S + \alpha U,
\end{equation}
where $\alpha$ represents the Costa's optimization parameter and a
cover-signal is modeled by a realizations set of Gaussian random
variables, independents and non stationary: $\mathcal{S} = \{S[1],
\ldots , S[G]\}$. According to the product rule
$$
    p(s | u, m) = \frac{ p(u | s, m) p_{S}(s)}{p(u | m)} \textrm ,
$$
we have:
\begin{equation}
    p(u | s, m) = \delta(u - Q_{\Delta}(s)) \textrm ,
\end{equation}
where $Q_{\Delta}(.)$ represents a scalar quantizer with step $\Delta$. In the other hand,
$$
    p(s | m) = \sum_{u} p(s | u, m) p(u | m)= \sum_{u} \delta(u - Q_{\Delta}(s)) p_S(s) \textrm .
$$
If we replace $S = \frac{X - \alpha U}{1 - \alpha}$ in the last equation, we obtain
$$
    p(x | m) = \frac 1 {1 - \alpha} \sum_{u} \delta \left(
        u - Q_{\Delta}\left(
            \frac{ x - \alpha u}{1 - \alpha}
        \right)
    \right) p_{S} \left(
        \frac{x - \alpha u}{1 - \alpha}
    \right) \textrm .
$$
When the information bits are equiprobable, we write:
$$
    p_{X}(x) = \frac 1 {2(1 - \alpha)} \sum_{u, m} \delta \left(
        u - Q_{\Delta} \left(
            \frac{ x - \alpha u}{1 - \alpha}
        \right)
    \right)
    p_{S} \left(
        \frac{x - \alpha u}{1 - \alpha}
    \right) \textrm .
$$

            \subsection{Demonstration of Eqn.~(\ref{eq:tcqlss})}\label{appendixB}

We note $\textbf{e}[i]$~--for $i = 1, \ldots , N$~-- the trellis
states and we suppose that all these states follow an uniform
distribution such as : $p_{E}(e) = 1/N$. In TCQ-based stego-system,
we substitute the cover-samples by $U_{(n,m,e)}$, $n \in
\mathcal{Z}$, the codeword of sub-codebook which corresponds to the
state $e$ and message-bit $m$. It is given by $U_{(n,
m,\textbf{e}[i])} = (n + m/2 - i/N) \Delta$ for $i = 1, \ldots, N/2$
and $U_{(n, m, \textbf{e}[i])} = U_{n, m, \textbf{e}[i - N/2]}$ for
$i = N/2 + 1, \ldots , N$. By leading on appendix \ref{appendixA},
the p.d.f. formulation of TCQ stego-signal for a fixed state $e$ is
:
\begin{eqnarray}
    && p(x | e) = \frac 1 {2 \left(
        1 - \alpha
    \right)} \sum_{n, m} 1_{ \left[
        - \frac 1 {2(1 - \alpha)}, \frac 1 {2(1 - \alpha)}
    \right]}(x - u_{(n, m, e)}) \nonumber \\ 
	&& \times p_{S} \left(
        \frac{x - \alpha u_{(n, m, e)}}{1 - \alpha}
    \right) \textrm ,
\end{eqnarray}
and
\begin{eqnarray}
    &&p_{X}(x)  =  \sum_{i = 1}^{N} p_{X}(x | \textbf{e}[i]) p_{E}(\textbf{e}[i]) \nonumber \\
    && =  \frac 1 {(1 - \alpha)} \sum_{n, m} \frac 1 N  \sum_{i = 1}^{N/2} 1_{ \left[
        - \frac 1 {2(1 - \alpha)}, \frac 1 {2(1 - \alpha)}
    \right]} \left(
        x - u_{(n, m, \textbf{e}[i])}
    \right) \nonumber \\
    && \times p_{S} \left(
        \frac{x - \alpha u_{(n, m, \textbf{e}[i])}}{1 - \alpha}
    \right) \textrm ,
\end{eqnarray}
if the number of states is large and by leading on the properties of
the Riemann sum, then:
\begin{eqnarray}
    p_{X}(x) & = & \frac 1 {1 - \alpha} \nonumber \\ && \sum_{n, m} \int_{0}^{\frac 1 2} 1_{\left[
        - \frac 1 {2(1 - \alpha)}, \frac 1 {2(1 - \alpha)}
    \right]} \left(
        x - (n + \frac m 2 - \gamma
    \right) \Delta) \nonumber \\
    && \times p_{S}\left(
        \frac{x - \alpha \left(
            n + \frac m 2 - \gamma
        \right) \Delta}{1 - \alpha}
    \right) \textrm {~d} \gamma \textrm .
\end{eqnarray}
If we replace $m$ by its two possible values, i.e. 0 or 1, and make
the following variable change $ Z = \frac{X - \alpha \gamma
\Delta}{1 - \alpha} $, we obtain:
$$
    p_{X}(x) = \frac 1 {
        \alpha \Delta
    } \int_{x - \frac{\alpha \Delta} 2}^{x + \frac{\alpha \Delta} 2} p_{S}(z) \textrm{~d} z =
    \frac 1 {\sigma_{w} \sqrt{12}} \int_{x - \sigma_{w} \sqrt{3}}^{x + \sigma_{w}\sqrt{3}} p_{S}(z) \textrm{~d} z \textrm .
$$

            \subsection{Demonstration of Eqn.~(\ref{eq_st})} \label{appendixC}

The transformation of the cover-signal is modeled by a realizations
set of Gaussian random variables, independents and non stationary,
i.e. $\mathcal{S}^{\scriptsize \textrm{st}} = \{S^{\scriptsize
\textrm{st}}[1], \ldots , S^{\scriptsize \textrm{st}}[G/\tau] \}$.
In addition, we take the spreading direction $\textbf{t}$ such as
$\forall i$, $\textbf{t}[i] = \pm \frac 1 {\sqrt \tau}$ and it is
modeled by a set of Gaussian, independents and non stationary random
variables, i.e. $\mathcal{T} = \{T[1], \ldots , T[N]\}$. Then, when
the ST-SCS is used to embed the message, the stego-signal $X$ is
given by $X = S + \alpha (U - S^{\scriptsize \textrm{st}}) T$, if we consider:
    $$
        S_{l}^{\scriptsize \textrm{st}} = \sum_{i = \tau l}^{\tau l + \tau - 1} S[i] \times T[i] =
        S[n] \times T[n] + \underbrace{\sum_{i \neq n} S[i] \times
        T[i]}_{Y_{n}[l]} \textrm ,
    $$
    where $Y$ is considered as a random variable modeled by a
    set
     $\mathcal{Y} = \{Y_{1}[1], \ldots , Y_{G}[G / \tau]\},$ then
\begin{equation}
    X = S + \alpha (U - S T - Y) T \textrm .
    \label{eq_ST_1}
\end{equation}
Since $\textbf{t}[i] = \pm 1 / \sqrt \tau$ and $\forall i$,
$\textbf{t}[i]^{2} = 1 / \tau$, thus the previous equations becomes
\begin{equation}
    X = \left(
        1 - \frac{\alpha}{\tau}
    \right) S - \alpha Y T + \alpha U T \textrm .
\end{equation}
Now, we compute the p.d.f of the codeword $U$ conditionally to
$S$, $Y$, $T$ and the message $m$:
\begin{equation}
    p(u | s, y, t, m) = \delta \left(
        u - Q_{\Delta} \left(
            s t + y
        \right)
    \right) \textrm ,
\end{equation}
where $\delta$ represents the Kronecker symbol. Therefore
\begin{equation}
    p(s | u, y, t, m) = \frac{\delta \left(
        u - Q_{\Delta} \left(
            s  t + y
        \right)
    \right) p(s | y, t, m)}{p(u | y, t, m)} \textrm .
\end{equation}
In this work, we consider $S$ as a random variable independent of
$T$ and  $Y$. Therefore $p(s | y, t, m) = p(s)$ and
\begin{equation}
    p(s | u, y, t, m) = \frac{\delta \left(
        u - Q_{\Delta} \left(
            s t + y
        \right)
    \right) p_{S}(s)}{p(u | y, t, m)} \textrm .
\end{equation}
Now, we make the following variable change:
\begin{equation}
    S = \frac{\tau}{\tau - \alpha} (X + \alpha   T - \alpha U T) \textrm .
\end{equation}
Then, we obtain
\begin{eqnarray}
     && p(x | u, y, t, m) =  \frac{\tau}{\tau - \alpha} \nonumber \\
    && \times \frac{\delta \left(
        u - Q_{\Delta}\left(
            \frac{\tau}{\tau - \alpha} \left(
                x + \alpha y t - \alpha u t
            \right)
        \right) t + y
    \right)}{p(u | y, t, m)} \nonumber \\
    && \times p_{S}\left(
        \frac{\tau}{\tau - \alpha} \left(
            x + y - \alpha u t
        \right)
    \right) \textrm ,
\end{eqnarray}
Since $T$ is a random variable which the realizations take just two
values $\pm 1 / \sqrt{\tau}$, and since $m$ is also considered as
equiprobable, the marginalization over this two variables and over
$U$ and $y$ gives:
\begin{eqnarray}
    && p_{X}(x)  =  \frac{\tau}{4 \left(
        \tau - \alpha
    \right)} \nonumber \\
    && \sum_{u, m, t} \int_{y} \delta \left(
        u - Q_{\Delta} \left(
            \frac{\tau}{\tau - \alpha} \left(
                x + \alpha y t - \alpha u  t
            \right) t + y
        \right)
    \right) \nonumber \\
    && \times p_{S} \left(
        \frac{\tau}{\tau - \alpha} \left(
            x + \alpha y t - \alpha u t
        \right)
    \right) p_{Y}(y) \textrm{~d} y \textrm .
\end{eqnarray}

\end{document}